\documentclass[graybox]{svmult}
\usepackage{type1cm,makeidx,graphicx,multicol,newtxtext}
\usepackage[bottom]{footmisc}
\usepackage[varvw]{newtxmath}

\makeindex            

\usepackage[doi=false,sorting=none,style=numeric-comp,giveninits=true,url=false,maxnames = 10]{biblatex}
\renewbibmacro{in:}{}
\DeclareFieldFormat[misc]{title}{``#1''\isdot}
\DeclareFieldFormat{journaltitle}{#1\isdot}
\DeclareFieldFormat[article,periodical]{volume}{\mkbibbold{#1}}
\DeclareFieldFormat{pages}{#1}
\AtEveryBibitem{%
  \clearfield{number}}

\usepackage{braket,bm}

\newcommand{\hilb}{\mathcal{H}}

\newcommand{\fock}{\mathcal{F}}
\newcommand{\focks}{\mathcal{F}}
\newcommand{\hfrak}{\mathfrak{H}}

\newcommand{\e}{\mathrm{e}}
\newcommand{\g}{\mathrm{g}}

\newcommand{\Psie}{\Psi_{\mathrm{e}}}
\newcommand{\Psig}{\Psi_{\mathrm{g}}}
\newcommand{\Phie}{\Phi_{\mathrm{e}}}
\newcommand{\Phig}{\Phi_{\mathrm{g}}}
\renewcommand{\a}[1]{a\!\left(#1\right)}
\newcommand{\adag}[1]{a^\dag\!\left(#1\right)}
\newcommand{\dOmega}{\mathrm{d}\Gamma(\omega)}

\begin{document}

\title*{Renormalization of spin--boson interactions mediated by singular form factors}
\author{Davide Lonigro}
\institute{Davide Lonigro \at Dipartimento di Fisica and MECENAS, Universit\`a di Bari, and INFN, Sezione di Bari, I-70126 Bari, Italy, EU, \email{davide.lonigro@ba.infn.it}}

\maketitle

\abstract{We study and discuss the extension of the rotating-wave spin--boson model, together with more general models describing a system--field coupling with a similar rotating-wave structure, to interactions mediated by possibly singular (non-normalizable) form factors satisfying a weaker growth constraint. To this purpose, a construction of annihilation and creation operators as continuous maps on a scale of Fock spaces, together with a rigorous renormalization procedure, is employed. }

\section{Introduction}
\label{sec:1}

The description of physical systems obeying the laws of quantum mechanics usually involves a distinction between a singled out subsystem under experimental control, usually finite-dimensional and simply referred to as an ``open quantum system'', and an external ``environment'' (or bath) which is typically infinite-dimensional and affects the system in a nontrivial way, yielding diverse effects like population decay, noise, or loss of quantum coherence~\cite{breuer2002theory,ingold2002path,grifoni1999dissipation,thorwart2004dynamics,clos2012quantification,costi2003entanglement}.

Two possible, and complementary, approaches to the study of such systems are available: one either focuses \textit{ab initio} on the open system alone, whose dynamics will be described by a completely positive and trace-preserving map on the Hilbert space of the open system (e.g. a Gorini--Kossakowski--Lindblad--Sudarshan semigroup~\cite{gorini1976completely,lindblad1976generators}), or tackles the problem directly at the ``system+environment'' level, which encompasses the choice of a legitimate self-adjoint operator---the \textit{Hamiltonian}---generating the joint unitary dynamics of the system and the environment; the evolution of the open system alone is thus recovered by tracing out the degrees of freedom of the environment. 

Spin--boson models and their many generalizations, modeling an open system (e.g. an atom) interacting with a structured boson field via annihilation and creation of field excitations, are typical examples of such models. They provide a treatable, yet usually realistic, description of many quantum phenomena of both fundamental and technological interest, and are widely used in diverse fields like quantum information, quantum optics and quantum simulation, to name a few~\cite{weiss2012quantum,leggett1987dynamics}. From the mathematical point of view, a lot of effort has been devoted to the investigation of the properties of spin--boson models~\cite{hubner1995spectral,hirokawa1999expression,hirokawa2001remarks,amann1991ground,davies1981symmetry,fannes1988equilibrium,hubner1995radiative,reker2020existence} and generalized spin--boson (GSB) models~\cite{arai1997existence,arai2000essential,takaesu2010generalized,falconi2015self,teranishi2015self,gsb}.

The prototypical example of such models is the following one. Let $\hilb=L^2_\mu(X)$ be the space of measurable complex-valued functions on a measure space $(X,\mu)$ with finite quadratic norm, that is,
\begin{equation}\label{eq:norm}
    \|f\|^2:=\int|f(k)|^2\;\mathrm{d}\mu(k)<\infty,
\end{equation}
where, here and in the following, all integrals are understood to be extended on the full space $X$. We denote as $\fock(\hilb)\equiv\fock$ the symmetric (or bosonic) Fock space with single-particle space $\hilb$, defined as usual:
\begin{equation}
    \fock=\bigoplus_{n\in\mathbb{N}}S_n\hilb^{(n)},\qquad \hilb^{(n)}=\bigotimes_{j=1}^n\hilb,
\end{equation}
with $S_n$ being the symmetrization operator on $\hilb^{(n)}$. This is the Hilbert space to be associated with the boson environment, with the Hilbert space of the atom simply being $\mathfrak{h}\simeq\mathbb{C}^2$. On their tensor product $\hfrak=\mathbb{C}^2\otimes\fock$, we define the following operator:
\begin{equation}
    H_0=\begin{pmatrix}
    \omega_\e&0\\0&\omega_\g
    \end{pmatrix}\otimes I+I\otimes\dOmega,
\end{equation}
with $\omega_\e,\omega_\g\in\mathbb{R}$, and $\dOmega$ being the second quantization of the multiplication operator associated with a real-valued, measurable function $\omega:X\rightarrow\mathbb{R}$; the latter is self-adjoint on its maximal domain $\mathcal{D}(\dOmega)$, see e.g.~\cite{reed1975fourier,bratteli1987operator}, and so is $H_0$ on $\mathcal{D}(H_0)=\mathbb{C}^2\otimes\mathcal{D}(\dOmega)$. The self-adjoint operator $H_0$ represents the free (or decoupled) Hamiltonian of the theory; the function $\omega$ is the \textit{dispersion relation} of the field. We shall restrict our attention to boson fields with strictly positive mass, that is, we shall suppose
\begin{equation}\label{eq:mass}
    m:=\inf_{k\in X}\omega(k)>0.
\end{equation}
We then define
\begin{equation}\label{eq:h}
    H_f=H_0+\lambda\left(\sigma^+\otimes\a{f}+\sigma^-\otimes\adag{f}\right),
\end{equation}
with $\lambda\in\mathbb{R}$ being a coupling constant, $\sigma^\pm$ being the usual ladder operators on $\mathbb{C}^2$:
\begin{equation}\label{eq:ladder}
    \sigma^+=\begin{pmatrix}
    0&1\\0&0
    \end{pmatrix}=(\sigma^-)^\dag,
\end{equation}
and $\a{f},\adag{f}$ being the annihilation and creation operators associated with some function $f\in\hilb$, which we denote as the \textit{form factor} of the model. This Hamiltonian, which is self-adjoint on $\mathcal{D}(H_f)=\mathcal{D}(H_0)$, corresponds to a variation of the standard spin--boson model in which counter-rotating terms---those not preserving the total number of excitations---are discarded, a procedure often denoted as the rotating-wave approximation (RWA)~\cite{agarwal1973rotating}, and finds applications in a wide range of situations; in particular, it reduces to the well-known Jaynes--Cummings model~\cite{shore1993jaynes,phoenix1991establishment,vogel1995nonlinear} in the case of a monochromatic field, simply obtained by taking $\mu$ as an atomic (Dirac) measure.

A limitation of the model is the following: in order for the expressions above to correctly define a self-adjoint operator on $\hfrak$, the form factor $f$ must be a square-integrable function, $f\in\hilb$. Without this constraint, the annihilation operator $\a{f}$, while still densely defined, fails to admit a densely defined adjoint~\cite{nelson1964interaction}. Nevertheless, a rigorous extension to \textit{singular} (i.e. non-normalizable, $f\notin\hilb$) form factors is both desirable for physical purposes, since singular form factors are often found in the physical literature, as well as expected on the basis of the comparison with similar models such as rank-one singular perturbations of differential operators~\cite{albeverio2000singular,simon1995spectral,posilicano2001krein} and Friedrichs (or Friedrichs--Lee) models~\cite{derezinski2002renormalization,facchi2021spectral,lonigro2021selfenergy}. It is also worth noting that, taking into account Eq.~\eqref{eq:mass}, this problem essentially corresponds to describing GSB models whose form factors exhibit ultraviolet (UV) divergencies; cf.~\cite{teufel2021hamiltonians,teufel2016avoiding,lampart2018particle,lampart2019nelson,binz2021abstract} for a detailed discussion of the same problem for other models of matter--field interaction.

An extension of the model above accommodating possibly singular form factors, together with analogous (but perturbative) results for all GSB models, was obtained in~\cite{gsb} by exploiting the formalism of Hilbert scales. Given $s\in\mathbb{R}$, define $\hilb_s$ as the space of all measurable functions satisfying the constraint
\begin{equation}
       \|\omega^{s/2}f\|^2=\int\omega(k)^s|f(k)|^2\;\mathrm{d}\mu<\infty;
\end{equation}
as a direct consequence of Eq.~\eqref{eq:mass}, this condition is either stronger (if $s>0$) or weaker (if $s<0$) than the normalization condition~\eqref{eq:norm}, whence a scale of Hilbert spaces,
\begin{equation}\label{eq:scale1}
    \ldots\supset\hilb_{-2}\supset\hilb_{-1}\supset\hilb_0\equiv\hilb\supset\hilb_1\supset\hilb_2\supset\ldots,
\end{equation}
each naturally endowed with the norm $\|f\|_{s}:=\|\omega^{s/2}f\|$, is obtained. In the same way, defining $\focks_s$ as the space of all sequences $\Psi=\{\Psi^{(n)}\}_{n\in\mathbb{N}}$ (with $\Psi^{(n)}$ being a completely symmetric $n$-particle function) satisfying
\begin{equation}
    \big\|(\dOmega+1)^{s/2}\Psi\big\|_\fock^2<\infty,
\end{equation}
with $\|\cdot\|_\fock$ denoting the norm on the Fock space $\focks$, a scale of Fock spaces is obtained:
\begin{equation}\label{eq:scale2}
    \ldots\supset\fock_{-2}\supset\fock_{-1}\supset\fock_0\equiv\fock\supset\fock_1\supset\fock_2\supset\ldots,
\end{equation}
each space being naturally endowed with the norm $\|\Psi\|_{\fock_s}:=\|(\dOmega+1)^{s/2}\Psi\|_\focks$. All inclusions in Eqs.~\eqref{eq:scale1} and~\eqref{eq:scale2} are dense.

With these definitions, the basic idea at the root of~\cite{gsb} is the following: given a singular form factor $f\in\hilb_{-s}$ for $s>0$, a legitimate self-adjoint operator to be associated with the (otherwise just formal) expression~\eqref{eq:h} can be constructed by interpreting $\a{f}$ and $\adag{f}$, instead as unbounded operators on the Fock space $\fock$, as continuous (bounded) maps on the scale of Fock spaces. The construction of such a ``singular'' spin--boson model was performed in the case $f\in\hilb_{-1}$ by means of a careful choice of the self-adjointness domain, and then further generalized to the case $f\in\hilb_{-2}$; remarkably, in the latter case one needs to ``trade'' the excitation energy $\omega_\e$ of the two-level system in Eq.~\eqref{eq:h} with a new parameter representing its \textit{dressed}, or renormalized, excitation energy. This is reminiscent of the renormalization procedures usually encountered in quantum field theories. The ``price to pay'' for this extension will be a modification of the original operator domain and, in the case $f\in\hilb_{-2}\setminus\hilb_{-1}$, even of its form domain.

In this work we shall summarize the aforementioned results for the spin--boson model with a rotating-wave approximation, with particular emphasis on the ideas at the ground of the renormalization procedure, and then discuss the extension of the same techniques to other instances of generalized spin--boson (GSB) models whose interaction term has an analogous rotating-wave structure, thus allowing for an explicit calculation of the resolvent instead of relying on perturbative methods. Namely, we shall investigate:
\begin{itemize}
    \item an extension of the rotating-wave spin--boson model in which the two-level system $\mathfrak{h}\simeq\mathbb{C}^2$ is replaced by a system of arbitrary (but finite) dimension decomposed into the direct sum of an ``excited'' and a ``ground'' sector, $\mathfrak{h}\simeq\mathfrak{h}_\e\oplus\mathfrak{h}_\g$;
     \item the multi-atom generalization of the rotating-wave spin--boson model, living on the Hilbert space $\hfrak=\mathbb{C}^{2^N}\otimes\fock$, with $N$ being the number of two-level systems.
 \end{itemize}
The work is organized as follows. In Section~\ref{sec:af} we briefly revise the formalism of singular annihilation and creation operators; in Section~\ref{sec:sb} we summarize the main results for the rotating-wave spin--boson model, first with form factor $f\in\hilb_{-1}$ and then with form factor up to $f\in\hilb_{-2}$; in Sections~\ref{sec:gsb}--\ref{sec:ibc} we investigate the aforementioned generalizations of said model, presenting preliminary results and discussing future developments; concluding remarks are provided in Section~\ref{sec:concl}.

\section{Singular annihilation and creation operators}\label{sec:af}

For convenience, we shall start by briefly recalling the construction of singular annihilation and creation operators presented in~\cite{gsb}, which enables us to model atom--field interactions mediated by singular form factors. 

Given a dispersion relation $\omega\geq m$ and the corresponding free field operator $\dOmega$, let $\{\hilb_s\}_{s\in\mathbb{R}}$, $\{\fock_s\}_{s\in\mathbb{R}}$ be the corresponding scales of Hilbert spaces as defined in the introduction. Recall that, in particular, for all $r,s\in\mathbb{R}$ the operator $(\dOmega+1)^{r}$ can be continuously extended to an isometry between the spaces $\fock_s$ and $\fock_{s-2r}$~\cite{simon1995spectral,albeverio2000singular,albeverio2007singularly}. Consequently, for all $s>0$ the triple $\left(\fock_s,\fock,\fock_{-s}\right)$ is a Gelfand triple~\cite{bohm1974rigged,de2005role,bohm1989dirac}, with the spaces $\focks_{\pm s}$ being mutually dual with respect to the duality pairing
\begin{equation}\label{eq:pairing}
    (\Psi,\Phi)\in\fock_{-s}\times\fock_{+s}\mapsto\left(\Psi,\Phi\right)_{\fock_{-s},\fock_s}:=\Braket{(\dOmega+1)^{-s/2}\Psi,(\dOmega+1)^{+s/2}\Phi}_\fock,
\end{equation}
with $\Braket{\cdot,\cdot}_\fock$ being the scalar product on the Fock space $\fock$.

\begin{proposition}[\cite{gsb}, Props. 3.4, 3.5 and 3.7]\label{prop:af}
Let $f\in\hilb_{-s}$ for some $s\geq1$. Then the following statements hold true:
\begin{itemize}
    \item[(i)] the operator $\a{f}:\fock_s\rightarrow\fock$ acting as $\a{f}\Omega=0$ and, for all $n\geq1$, as
    \begin{equation}\label{eq:af}
    \left(\a{f}\Psi^{(n)}\right)(k_1,\dots,k_{n-1})=\sqrt{n}\int\overline{f(k_n)}\Psi^{(n)}(k_1,\dots,k_{n-1},k_n)\,\mathrm{d}\mu(k_n),
    \end{equation}
    is well-defined and bounded;
    \item[(ii)]\; its adjoint $\adag{f}:=\a{f}^*:\fock\rightarrow\fock_{-s}$ with respect to the pairing~\eqref{eq:pairing} is likewise well-defined and bounded, and, for $\Psi\in\fock_{+1}$,
    \begin{eqnarray}\label{eq:adagf}
    \left(\adag{f}\Psi^{(n)}\right)(k_1,...,k_{n},k_{n+1})&=&\frac{1}{\sqrt{n+1}}\bigg(\sum_{j=1}^n\Psi^{(n)}(k_1,...,\overbrace{k_{n+1}}^{j\text{th}},...,k_n)f(k_j)\nonumber\\&&+\Psi^{(n)}(k_1,...,k_n)f(k_{n+1})\bigg);
    \end{eqnarray}
    \item [(iii)]\;\; finally, there exists a sequence of regular form factors $\{f^i\}_{i\in\mathbb{N}}\subset\hilb$ such that
\begin{equation}
	\lim_{i\to\infty}\left\|a(f)-a(f^i)\right\|_{\mathcal{B}(\focks_s,\focks)}=0,\qquad 	\lim_{i\to\infty}\left\|a^\dag(f)-a^\dag(f^i)\right\|_{\mathcal{B}(\focks,\focks_{-s})}=0,
\end{equation}
and this happens if and only if $\|f-f^i\|_{-s}\to0$.
    \end{itemize}
\end{proposition}
Above, $\Omega$ is the vacuum state of the boson field, i.e. the single normalized element (up to an irrelevant phase term) of $\hilb^{(0)}$, while $\mathcal{B}(\cdot,\cdot)$ stands for the linear space of continuous maps between two Banach (in this case Hilbert) spaces. 

The proof of (i)--(ii) is basically an application of standard estimates on the norm of $\a{f}\Psi$ (cf.~\cite[Eq. (5)]{nelson1964interaction} or~\cite{ginibre2006partially}) translated within the language of Fock scales, while (iii) follows from the fact that $\hilb$ is densely embedded into $\hilb_{-s}$. Clearly, when $f\in\hilb$, the action of both operators above coincide with that of the ``standard'' annihilation and creation operators (defined as unbounded operators on $\fock$) whenever the latter are defined, thus justifying our notation and nomenclature; when $f\in\hilb_{-s}\setminus\hilb$, i.e. in the case of non-normalizable form factors, the singular operators can be approximated by the regular ones in a suitable topology.

\section{Renormalization of the rotating-wave spin--boson model}\label{sec:sb}

Let us start by considering the spin--boson model in Eq.~\eqref{eq:h} with a regular form factor $f\in\hilb$; without loss of generality, we shall set the ground energy of the atom to zero, $\omega_\g=0$. By using the obvious isomorphism $\hfrak=\mathbb{C}^2\otimes\fock\simeq\fock\oplus\fock$, the model can be conveniently written in a formal matrix fashion:
\begin{equation}\label{eq:rwa}
    H_{f,\omega_\e}=\begin{pmatrix}
    \omega_\e+\dOmega&\lambda\a{f}\\
    \lambda\adag{f}&\dOmega
    \end{pmatrix},
\end{equation}
with domain $\mathcal{D}\!\left(H_{f,\omega_\e}\right)=\mathbb{C}^2\otimes\mathcal{D}(\dOmega)\simeq\mathcal{D}(\dOmega)\oplus\mathcal{D}(\dOmega)$. For future convenience, in the remainder of this section we will explicitly indicate the dependence on $\omega_\e$. The domain includes, in particular, the state
\begin{equation}\label{eq:psi0}
    \Psi_0=\begin{pmatrix}
    \Omega\\0
    \end{pmatrix},
\end{equation}
representing the configuration in which the atom is in its excited state and the field is in its vacuum state $\Omega\in\fock$. Notice that, in particular, the mean value and the variance of the total energy of the system in this state read
\begin{eqnarray}\label{eq:variance}
    \Braket{H_{f,\omega_\e}}_{\Psi_0}&=&\omega_\e,\\ \Braket{H_{f,\omega_\e}^2}_{\Psi_0}-\Braket{H_{f,\omega_\e}}^2_{\Psi_0}&=&\lambda^2\|f\|^2.
\end{eqnarray}

If $f\notin\hilb$, Eq.~\eqref{eq:rwa} does not define a legitimate operator on $\fock$ with the domain prescription chosen above: even by interpreting $\a{f},\adag{f}$ in the sense of Prop.~\ref{prop:af}, the latter operator has values outside the Fock space $\fock$ when applied on $\mathcal{D}(\dOmega)\oplus\mathcal{D}(\dOmega)$. A different domain prescription, which will turn out to depend both on the choice of form factor $f$ as well as the coupling constant $\lambda$, is therefore required. We shall present the main results hereafter; a summary of our findings can be found in Table~\ref{tab:2} at the end of this section.

\subsection{Case $f\in\hilb_{-1}$}
Let us start by addressing the case $f\in\hilb_{-1}$. First, to this purpose it is useful to compute the resolvent of the regular model above. The latter reads as follows~\cite{gsb}: for all $\Psie,\Psig\in\fock$, and $z\in\mathbb{C}\setminus\mathbb{R}$,
\begin{equation}\label{eq:ressing}
	\frac{1}{H_{f\!,\omega_\e}-z}\!\begin{pmatrix}
	\Psie\\\Psig
	\end{pmatrix}=\begin{pmatrix}
	\mathcal{G}_{f,\omega_\e}^{-1}(z)\Bigl(\Psie-\lambda\a{f}\frac{1}{\dOmega-z}\Psig\Bigr)\\
	\frac{1}{\dOmega-z}\Psig\!-\!\lambda\frac{1}{\dOmega-z}\adag{f}\mathcal{G}^{-1}_{f,\omega_\e}(z)\Bigl(\Psie-\lambda\a{f}\frac{1}{\dOmega-z}\Psig\Bigr)
	\end{pmatrix}
\end{equation}
where $\mathcal{G}^{-1}_{f,\omega_e}(z)$, the \textit{propagator} of the model, is defined as the bounded inverse~\cite[Lemma 5.3]{gsb} of the operator with domain $\mathcal{D}(\dOmega)$
\begin{equation}\label{eq:propagator}
    \mathcal{G}_{f,\omega_\e}(z)=\omega_\e-z+\dOmega-\lambda^2\mathcal{S}_f(z),
\end{equation}
where
\begin{equation}\label{eq:sf}
 \mathcal{S}_f(z)=\a{f}\frac{1}{\dOmega-z}\adag{f}.
\end{equation}
Now, a close scrutiny to Eqs.~\eqref{eq:ressing}--\eqref{eq:sf} shows that, provided that one interprets $\a{f}$ and $\adag{f}$ as continuous maps on the scale of Fock spaces constructed before (see Prop.~\ref{prop:af}), the latter expression is indeed well-defined even when, more generally, $f\in\hilb_{-1}$; indeed, 
\begin{equation}\label{eq:scheme}
    \fock\xrightarrow{\adag{f}}\fock_{-1}\xrightarrow{(\dOmega-z)^{-1}}\fock_{+1}\xrightarrow{\a{f}}\fock,
\end{equation}
whence the map $\mathcal{S}_f(z)$ in Eq.~\eqref{eq:sf} is bounded on $\fock$. This simple observation is at the root of the following theorem:
\begin{theorem}[\cite{gsb}, Theorem 5.5]\label{thm:1}
Let $f\in\hilb_{-1}$, $\omega_\e,\lambda\in\mathbb{R}$, and let $H_{f,\omega_\e}$ be the operator on $\fock\oplus\fock$ with domain
\begin{equation}\label{eq:singdom}
	\mathcal{D}(H_{f,\omega_\e})=\left\{
	\begin{pmatrix}
	\Phie\\\Phig-\lambda\frac{1}{\dOmega+1}\adag{f}\Phie
	\end{pmatrix}:\;\Phie,\Phig\in\mathcal{D}\!\left(\dOmega\right)
	\right\},
\end{equation}
acting as
\begin{equation}\label{eq:action}
	H_{f,\omega_\e}\begin{pmatrix}
	\Phie\\\Phig-\lambda\frac{1}{\dOmega+1}\adag{f}\Phie
	\end{pmatrix}=\begin{pmatrix}
	\left(\omega_{\mathrm{e}}+\dOmega-\lambda^2\mathcal{S}_f(-1)\right)\Phie+\lambda\,\a{f}\Phig\\
	\dOmega\Phig+\lambda\frac{1}{\dOmega+1}\adag{f}\Phie
	\end{pmatrix}.
\end{equation}
Then:
\begin{itemize}
    \item[(i)] \, for $f\in\hilb$, the operator coincides with the spin--boson Hamiltonian in Eq.~\eqref{eq:rwa};
    \item[(ii)]\,\,for $f\in\hilb_{-1}\setminus\hilb$, the operator is self-adjoint, its resolvent reads as in Eq.~\eqref{eq:ressing}, and there exists a sequence $\{f^i\}_{i\in\mathbb{N}}\subset\hilb$ such that $H_{f^i,\omega_\e}\to H_{f,\omega_\e}$ in the norm resolvent sense. In particular, this happens iff
    \begin{equation}\label{eq:singlimit}
        \|f^i-f\|_{-1}\to0.
    \end{equation}
\end{itemize}
\end{theorem}
Points (i) and the first part of (ii) are essentially a direct consequence of the mapping properties of $\a{f}$ and $\adag{f}$, while the last part of (ii) follows from Prop.~\ref{prop:af}(iii).

It is worth discussing the structure of the operator domain~\eqref{eq:singdom}, which involves the presence of an additional term $\lambda(\dOmega+1)^{-1}\adag{f}$ in the second component of the states. If $f\in\hilb$, the latter term is in $\mathcal{D}(\dOmega)$ and can be simply reabsorbed in the definition of $\Phi_\g$, so that the excited and ground state wavefunctions can be chosen independently from each other. Instead, if $f\in\hilb_{-1}\setminus\hilb$, the additional term does not belong to $\mathcal{D}(\dOmega)$ and cannot be reabsorbed, thus behaving as a ``singular'' potential effectively coupling the two components. Physically, in order for the variance of the total energy of the system in a state to be finite, its free boson energy must diverge. In this sense, in the limit $H_{f^i,\omega_\e}\to H_{f,\omega_\e}$, the operator domain of the model exhibits a highly ``irregular'' behavior---it is constant for all $i\in\mathbb{N}$, and changes abruptly in the limit---while the resolvent behaves regularly.

Finally, notice that the operator domain in the case $f\in\hilb_{-1}\setminus\hilb$ does \textit{not} include $\Psi_0$, the excited state of the atom interacting with the boson vacuum as defined in Eq.~\eqref{eq:psi0}; however, $\Psi_0$ is still in the form domain. This is in full accordance with Eq.~\eqref{eq:variance}: the variance of the total energy must diverge, while its average value does not change.

\subsection{Case $f\in\hilb_{-2}$}

Without any further modification, the result above cannot be extended to a larger class of form factors. Namely, if $f\in\hilb_{-s}\setminus\hilb_{-1}$ for some $s>1$, then the operator $\mathcal{S}_f(z)$ in Eq.~\eqref{eq:sf} (and thus the propagator itself) fails to be well-defined. Indeed, again by the properties of $\a{f}$ and $\adag{f}$ collected in Prop.~\ref{prop:af}, we have
\begin{equation}\label{eq:mappings1}
\fock\xrightarrow{\adag{f}}\fock_{-s}\xrightarrow{(\dOmega-z)^{-1}}\fock_{-s+2}\supset\fock_{+s},
\end{equation}
the latter inclusion being proper since $s>1$, whence $\a{f}$ cannot be applied.

However, as long as $s\leq2$, the \textit{difference} between the values of $\mathcal{S}_f(z)$ in two distinct points outside the spectrum of $\dOmega$, say $z\in\mathbb{C}\setminus\mathbb{R}$ and $z_0=-1$, is in fact well-defined. Indeed, because of the first resolvent identity, the difference $(\dOmega-z)^{-1}-(\dOmega+1)^{-1}$ is ``more regular'' than each of the two terms alone, i.e. it satisfies
\begin{equation}
(\dOmega-z)^{-1}-(\dOmega+1)^{-1}:\focks_{-s}\rightarrow\focks_{-s+4},
\end{equation}
whence
\begin{equation}\label{eq:mappings2}
\fock\xrightarrow{\adag{f}}\fock_{-s}\xrightarrow{(\dOmega-z)^{-1}-(\dOmega+1)^{-1}}\fock_{-s+4}\subset\fock_{+s}\xrightarrow{\a{f}}\fock,
\end{equation}
the latter inclusion holding provided that $s\leq2$.

This observation suggests that it may be possible to define a \textit{renormalized} version of the operator-valued map $\mathcal{S}_f(z)$ accommodating a form factor $f\in\hilb_{-s}\setminus\hilb_{-1}$ up to $s=2$. To this purpose, however, a slightly stronger assumption will be needed. Given $s\in(1,2]$ and $r\in[s-1,1]$, we define
\begin{equation}\label{eq:quick}
    \hilb_{-s}^r:=\left\{
    f\in\hilb_{-s}:\,\int\frac{|f(k)|^2}{[\omega(k)+(n-1)m]^s}\,\mathrm{d}\mu(k)=\mathcal{O}(n^{s-r})
    \right\},
\end{equation}
that is, the subspace of functions in $\hilb_{-s}$ such that, in addition, the sequence of integrals in Eq.~\eqref{eq:quick} decays ``sufficiently quickly'' as $n\to\infty$. As shown in~\cite[Lemma 6.2 and Lemma 6.5]{gsb}, this technical request, which is typically satisfied in realistic models, allows one to define a renormalized propagator in the following sense: there exists an (unbounded) operator-valued map $\tilde{\mathcal{S}}_f(z)$ such that
\begin{itemize}
    \item if $f\in\hilb_{-1}$, then $\tilde{\mathcal{S}}_f(z)=\mathcal{S}_f(z)+\|f\|^2_{-1}$;
    \item if $f\in\hilb_{-s}^r$ for some $s\in(1,2]$ and $r\in[s-1,1]$, then $\tilde{\mathcal{S}}_f(z)$ is still a well-defined operator which, in addition, is relatively bounded with respect to $\dOmega$ (and even \textit{infinitesimally} relatively bounded if $r<1$).
\end{itemize}
Consequently, looking at the definition~\eqref{eq:propagator} of the operator $\mathcal{G}_{f,\omega_\e}(z)$, the following holds: given $\tilde{\omega}_\e\in\mathbb{R}$ and the operator on $\fock$ defined via
\begin{equation}\label{eq:propagator_renorm}
\tilde{\mathcal{G}}_{f,\tilde{\omega}_\e}(z)=\tilde{\omega}_\e-z+\dOmega-\lambda^2\tilde{\mathcal{S}}_f(z),
\end{equation}
we can conclude that
\begin{itemize}
    \item if $f\in\hilb_{-1}$, then $\tilde{\mathcal{G}}_{f,\tilde{\omega}_\e}(z)=\mathcal{G}_{f,\omega_\e}(z)$ with the latter as defined in Eq.~\eqref{eq:propagator}, provided that
    \begin{equation}\label{eq:shift}
        \tilde{\omega}_\e=\omega_\e+\lambda^2\|f\|^2_{-1};
    \end{equation}
    \item if $f\in\hilb_{-s}^r$ for some $s\in(1,2]$ and $r\in[s-1,1]$, then $\tilde{\mathcal{G}}_{f,\tilde{\omega}_\e}(z)$ is still a well-defined operator with domain $\mathcal{D}(\dOmega)$.
\end{itemize}
Heuristically, this can be interpreted as follows: the divergence of $\mathcal{S}_f(z)$ for $f\notin\hilb_{-1}$ is ``cured'' by adding an ``infinite constant'' to it; correspondingly, the same constant must be added to the excitation energy $\omega_\e$ of the atom. 

This construction allows us to prove the following theorem, which further extends the results of Theorem~\ref{thm:1}:
\begin{theorem}[\cite{gsb}, Theorem 6.6]\label{thm:2}
Let $f\in\hilb^r_{-s}$ for some $s\in[1,2]$ and $r\in[s-1,1]$, $\tilde{\omega}_\e,\lambda\in\mathbb{R}$, and let $\tilde{H}_{f,\tilde{\omega}_\e}$ be the operator on $\fock\oplus\fock$ with domain as in Eq.~\eqref{eq:singdom},
acting as
\begin{equation}\label{eq:action2}
	\tilde{H}_{f,\tilde{\omega}_\e}\begin{pmatrix}
	\Phie\\\Phig-\lambda\frac{1}{\dOmega+1}\adag{f}\Phie
	\end{pmatrix}=\begin{pmatrix}
	\left(\tilde{\omega}_{\mathrm{e}}+\dOmega-\lambda^2\tilde{\mathcal{S}}_f(-1)\right)\Phie+\lambda\,\a{f}\Phig\\
	\dOmega\Phig+\lambda\frac{1}{\dOmega+1}\adag{f}\Phie
	\end{pmatrix}.
\end{equation}
Then
\begin{itemize}
    \item[(i)] \,if $s=1$, the operator coincides with the spin--boson Hamiltonian in Eq.~\eqref{eq:rwa}, provided that $\omega_\e$ is as given by Eq.~\eqref{eq:shift};
    \item[(ii)]\,\,if $s\in(1,2]$ and $r\in[s-1,1)$, then, for all $\lambda\in\mathbb{R}$, the operator is self-adjoint, and there exist sequences $\{f^i\}_{i\in\mathbb{N}}\subset\hilb$, $\{\omega^i_\e\}_{i\in\mathbb{N}}\subset\mathbb{R}$ such that $H_{f^i,\omega_\e^i}\to\tilde{H}_{f,\tilde{\omega}_\e}$ in the strong resolvent sense; in particular, this happens iff
    \begin{equation}\label{eq:singlimit2}
        \|f^i-f\|_{-s}\to0\qquad\text{and}\qquad\omega^i_\e+\lambda^2\|f^i\|^2_{-1}\to\tilde{\omega}_\e;
    \end{equation}
    \item[(iii)]\,\,\,\,if $s\in(1,2]$ and $r=1$, then the same statements as in point (ii) hold for sufficiently small $\lambda$.
    \end{itemize}
\end{theorem}
The proof follows similar arguments as the one of Theorem~\ref{thm:1}. Eq.~\eqref{eq:singlimit2} accounts for the interpretation of the parameter $\tilde{\omega}_\e$ as the \textit{renormalized} (or dressed) excitation energy of the atom: since $f\notin\hilb_{-1}$, necessarily $\|f^i\|_{-1}$ diverges and thus, in order for the latter to be finite, and whence the model to be well-defined, the \textit{bare} one must diverge as well. This is a simple example of a renormalization procedure. It is not required in the case $f\in\hilb_{-1}$, where both energies are finite and only differ by a shift. 

We point out that, while in the case $f\in\hilb_{-1}$ Theorem~\ref{thm:1} ensures the possibility to approximate the singular model via regular one in the \textit{norm} resolvent sense, in the more general case $f\in\hilb_{-s}^r$ Theorem~\ref{thm:2} only provides \textit{strong} resolvent sense. Mathematically, this is essentially a consequence of the fact that, differently from what happens in the case $f\in\hilb_{-1}$, here the operator $\tilde{\mathcal{S}}_f(z)$ is unbounded. Whether this result may be improved, at least under additional assumptions, is a currently open question.

Finally, as a further difference from the case $f\in\hilb_{-1}$, notice that now the state $\Psi_0$ does not even belong to the form domain of the renormalized operator, again in full accordance with Eq.~\eqref{eq:variance}: since the bare excitation energy ``is infinite'' (in the sense discussed before), necessarily the average value of the total energy in that state must diverge as well. A summary of all possible cases, as given by Theorems~\ref{thm:1}--\ref{thm:2}, is reported in Table~\ref{tab:2}.

\begin{table}[!t]
\caption{Classification of the main properties of the regular spon--boson model and the singular one for all values $1\leq s\leq2$ and $s-1\leq r\leq 1$, as given by Theorems~\ref{thm:1}--\ref{thm:2}. In order: type of approximation by regular models, admissible values of the coupling constant, average value and variance of the total energy of the system in the state $\Psi_0$. Notice that the finiteness (or lack thereof) of the latter two quantities reflects whether the form domain and operator domain, respectively, of the corresponding singular model are dependent or independent of the coupling.}
\label{tab:2}
\begin{tabular}{llccc}
\hline\noalign{\smallskip}
Coupling class & Approxim. & Coupling const. & \;Mean energy $\Psi_0$\; & Variance energy $\Psi_0$\; \\
\noalign{\smallskip}\svhline\noalign{\smallskip}
$\hilb$ &   & Arbitrary & $\omega_\e$ & $\lambda^2\|f\|^2$\\
$\hilb_{-1}\setminus\hilb$ & Norm resolv. & Arbitrary & $\omega_\e$ & $\infty$ \\
$\hilb_{-s}^r\setminus\hilb_{-1}$, \,$s>1,r\!<1\,$ & Strong resolv. & Arbitrary & $\infty$ & $\infty$\\
$\hilb_{-s}^r\setminus\hilb_{-1}$, \,$s>1,r=1\,$ & Strong resolv. & Small & $\infty$ & $\infty$\\
\noalign{\smallskip}\hline\noalign{\smallskip}
\end{tabular}
\end{table}

\section{Renormalization of higher-dimensional models}\label{sec:gsb}

The construction presented above is not exclusive of the qubit scenario: similar results as in the two-level case may be obtained, \textit{mutatis mutandis}, to some classes of generalized spin--boson (GSB) models whose interaction exhibits a similar rotating-wave structure. We shall hereafter examine some preliminary results for a higher-dimensional generalization of the model above.

We shall replace the two-level atom of Section~\ref{sec:sb} with a multi-level atom, associated with a Hilbert space $\mathfrak{h}$ which can be decomposed into the direct sum of two sectors, $\mathfrak{h}\simeq\mathfrak{h}_\e\oplus\mathfrak{h}_\g$, effectively behaving like the two levels of a spin--boson model, in the following sense. Choosing the decoupled Hamiltonian $H_0$ on the Hilbert space $\hfrak=\mathfrak{h}\otimes\fock$ as
\begin{equation}
    H_0=\begin{pmatrix}
    E_\e&0\\0&E_\g
    \end{pmatrix}\otimes I+I\otimes\dOmega,
\end{equation}
with $E_\e\in\mathcal{B}(\mathfrak{h}_\e)$ and $E_\g\in\mathcal{B}(\mathfrak{h}_\g)$ symmetric and nonnegative, we set, for some integer $r$,
\begin{equation}\label{eq:gsb1}
     H=H_0+\lambda\sum_{j=1}^r\left(\Sigma^+_j\otimes\a{f_j}+\Sigma^-_j\otimes\adag{f_j}\right),
     \end{equation}
where $\Sigma_j^+\in\mathcal{B}(\mathfrak{h}_\g,\mathfrak{h}_\e)$, $\Sigma_j^-=(\Sigma_j^-)^\dag\in\mathcal{B}(\mathfrak{h}_\e,\mathfrak{h}_\g)$, and $f_1,\dots,f_r\in\hilb$ are form factors. 

The physical interpretation of this model is straightforward. The system behaves like a multilevel atom whose energy levels can be grouped into two sectors, $\mathfrak{h}_\e$ and $\mathfrak{h}_\g$; the interaction with the boson field causes $r$ possible transitions between levels belonging to separate sectors, while transitions between levels of the same sector are not allowed; the $j$th allowed transition from $\mathfrak{h}_\e$ to $\mathfrak{h}_\g$ (resp. from $\mathfrak{h}_\g$ to $\mathfrak{h}_\e$) involves the creation (resp. annihilation) of a boson with wavefunction $f_j$. In the case $\mathfrak{h}_\g\simeq\mathbb{C}$ (i.e. a single ground level), this model was first studied in~\cite{garraway1997nonperturbative}, and the properties of the corresponding reduced dynamics were investigated in~\cite{lonigro2022beyond,chruscinski2022markovianity}.

Following a similar strategy as in the spin--boson case, we can exploit the isomorphism $\hfrak=\mathfrak{h}\otimes\fock\simeq\hfrak_\e\oplus\hfrak_\g$, with $\hfrak_s:=\mathfrak{h}_s\otimes\fock$ for $s\in\{\e,\g\}$, and write the model above in a formal matrix fashion:
\begin{equation}\label{eq:rwa_gsb1}
    H_{\bm{f}}=\begin{pmatrix}
    h_\e&\lambda A_{\bm{f}}\\
    \lambda A^\dag_{\bm{f}}& h_\g
    \end{pmatrix},
\end{equation}
where we are using the shorthands $\bm{f}=(f_1,\dots,f_r)$,
\begin{equation}
    h_s=E_s\otimes I+I\otimes\dOmega,\qquad s\in\{\e,\g\}
\end{equation}
and
\begin{equation}
    A_{\bm{f}}=\sum_{j=1}^r\Sigma^+_j\otimes\a{f_j}.
\end{equation}
Following the clear analogies with the spin--boson model analyzed in Section~\ref{sec:sb}, the extension of this model to the case $f_1,\dots,f_r\in\hilb_{-1}$ can be performed by following a similar route as in Theorem~\ref{thm:1}. Indeed, in analogy with Eq.~\eqref{eq:propagator}, define
\begin{equation}\label{eq:propagator_gsb}
    \mathcal{G}_{\bm{f}}(z)=h_\e-z-\lambda^2\mathcal{S}_{\bm{f}}(z),
\end{equation}
where
\begin{equation}
    \mathcal{S}_{\bm{f}}(z)=A_{\bm{f}}\frac{1}{h_\g-z}A^\dag_{\bm{f}}=\sum_{j,\ell=1}^r\Sigma_j^+\otimes\a{f_j}\frac{1}{h_\g-z}\,\Sigma_j^-\otimes\adag{f_\ell},
\end{equation}
the latter being still a well-defined, bounded operator on $\hfrak_\e$ as long as $f_1,\dots,f_r\in\hilb_{-1}$, because of the usual mapping properties of singular annihilation and creation operators. Remarkably, in the particular case $\mathfrak{h}_\g\simeq\mathbb{C}$ (i.e. the ground sector being one-dimensional), in which necessarily $\Sigma^+_j=\ket{\e_j}$ and $\Sigma^-_j=\bra{\e_j}$ for some family of vectors $\{\e_j\}_j\subset\mathfrak{h}_\e$, the latter expression simplifies as follows:
\begin{equation}\label{eq:onedim}
    \mathcal{S}_{\bm{f}}(z)=\sum_{j,\ell=1}^r\ket{\e_j}\!\!\bra{\e_\ell}\otimes\a{f_j}\frac{1}{\omega_\g+\dOmega-z}\adag{f_\ell}\nonumber,
\end{equation}
which is clearly akin to Eq.~\eqref{eq:sf}.

With these definitions, in analogy with the construction presented in Theorem~\ref{thm:1}, one can readily introduce a singular version of the model accommodating arbitrary form factors $f_1,\dots,f_r\in\hilb_{-1}$. The singular model has domain
\begin{equation}
    \mathcal{D}\left(H_{\bm{f}}\right)=
\left\{
	\begin{pmatrix}
	\Phie\\\Phig-\lambda\frac{1}{h_\g+1}A^\dag_{\bm{f}}\Phie
	\end{pmatrix}:\;\Phie,\in\mathcal{D}(h_\e),\,\Phig\in\mathcal{D}(h_\g),
	\right\},
\end{equation}
its action on the latter being given by
\begin{equation}\label{eq:action_gsb}
	H_{\bm{f}}\begin{pmatrix}
	\Phie\\\Phig-\lambda\frac{1}{h_\g+1}A^\dag_{\bm{f}}\Phie
	\end{pmatrix}=\begin{pmatrix}
	\left(h_\e-\lambda^2\mathcal{S}_{\bm{f}}(-1)\right)\Phie+\lambda\,A_{\bm{f}}\Phig\\
	h_\g\Phig+\lambda\frac{1}{h_\g+1}A^\dag_{\bm{f}}\Phie
	\end{pmatrix},
\end{equation}
and its resolvent being given by an expression analogous to Eq.~\eqref{eq:ressing}, with $\mathcal{G}_{\bm{f}}(z)$ as in Eq.~\eqref{eq:propagator_gsb}. A direct check shows that, as in the spin--boson case, the model reduces to the regular one whenever all form factors are normalizable, $f_1,\dots,f_r\in\hilb$, and can be approximated in the norm resolvent sense by sequences of regular ones; for the latter purpose, given $f_1,\dots,f_r\in\hilb_{-1}\setminus\hilb$, it suffices to take into account sequences $\{f^i_1\}_{i\in\mathbb{N}},\dots,\{f^i_r\}_{i\in\mathbb{N}}\subset\hilb$ such that $\|f^i_1-f_1\|_{-1}\to0,\dots,\|f^i_r-f_r\|_{-1}\to0$ as $i\to\infty$, such sequences always existing since $\hilb$ is densely embedded into $\hilb_{-1}$.

As in the spin--boson case, such results cannot be directly extended to the case of form factors $f_1,\dots,f_r\in\hilb_{-s}\setminus\hilb_{-1}$ for $s>1$, since the propagator fails to be defined in such a case, cf.~Eq.~\eqref{eq:mappings1}; however, again as a consequence of the first resolvent identity, the \textit{difference} $\mathcal{S}_{\bm{f}}(z)-\mathcal{S}_{\bm{f}}(-1)$ is indeed well-defined, cf.~Eq.~\eqref{eq:mappings2}, thus suggesting the possibility of pursuing a renormalization procedure analogous to (but mathematically more intricate than) the one for the spin--boson model. We shall explore this possibility elsewhere.

\section{Perspective: many-body spin--boson model}\label{sec:multi}

Let us now consider the many-body generalization of the spin--boson model. The corresponding Hamiltonian, to be defined on the Hilbert space $\hfrak=\mathbb{C}^{2^N}\otimes\fock$, can be obtained as follows. The decoupled Hamiltonian reads
\begin{equation}
	H_0=K\otimes I+I\otimes\dOmega,
\end{equation}
where
\begin{equation}\label{eq:freesbn}
	K=\sum_{j=1}^N\left( I\otimes\dots\otimes K_j\otimes\dots\otimes I\right),\qquad K_j=\begin{pmatrix}
		\omega_{\mathrm{e},j}&0\\0&\omega_{\mathrm{g},j}
	\end{pmatrix},
\end{equation}
with $\omega_{\e,j}$, $\omega_{\mathrm{g},j}\in\mathbb{R}$. Given $f_1,\dots,f_N\in\hilb$, we set
\begin{equation}\label{eq:multiatom}
	H_{\bm{f}}=H_0+\lambda\sum_{j=1}^N\left(\sigma^+_j\otimes\a{f_j}+\sigma^-_j\otimes\adag{f_j}\right),
\end{equation}
where
\begin{equation}
	\sigma^\pm_j=\bigoplus_{j=1}^N\bigl( I\otimes\dots\otimes \overbrace{\sigma^\pm}^{j\text{th}}\otimes\dots\otimes I\bigr).
\end{equation}
Leaving to a future work a detailed discussion about the implementation of singular form factors in this model, we shall show that, as a starting point, this Hamiltonian can be indeed recast in a matrix fashion akin to the one in Eq.~\eqref{eq:rwa} for the spin--boson model, or Eq.~\eqref{eq:rwa_gsb1} for its multilevel generalization, thus allowing for similar calculations. 

To this purpose, since the dimension of the spin subsystem grows exponentially with $N$, it will be largely convenient to introduce a clever representation by packing together all states in which an equal number of atoms are, say, in their excited state:
\begin{equation}\label{eq:decomp}
	\hfrak\simeq\bigoplus_{j=0}^N\hfrak^{(j)},\qquad\hfrak^{(j)}=\bigoplus_{\ell=1}^{n_j}\fock,\qquad n_j=\binom{N}{j}
\end{equation}
with $\hfrak^{(j)}$ including the boson wavefunctions corresponding to the $n_j$ states in which $j$ atoms are excited and $N-j$ atoms are in their ground state. Notably, the number of such sectors grows linearly with $N$. Similarly, for all $s\in\mathbb{R}$ we will write
\begin{equation}\label{eq:decomp_s}
	\hfrak_{s}\simeq\bigoplus_{j=0}^N\hfrak^{(j)}_{s},\qquad\hfrak^{(j)}_{s}=\bigoplus_{\ell=1}^{n_j}\fock_{s}.
\end{equation}
We shall briefly examine explicitly the case $N=2$. In this case, the decomposition~\eqref{eq:decomp} reads $\hfrak\simeq\hfrak^{(0)}\oplus\hfrak^{(1)}\oplus\hfrak^{(2)}$, with $\hfrak^{(0)}=\hfrak^{(2)}=\fock$ and $\hfrak^{(1)}=\fock\oplus\fock$. Correspondingly, the Hamiltonian $H_{f_1,f_2}$ can be written as
\begin{equation}\label{eq:twoatom2}
	H_{f_1,f_2}=\left(\begin{array}{c|cc|c}
		h_{\mathrm{ee}}&\lambda\a{f_2}&\lambda\a{f_1}&0\\\hline
		\lambda\adag{f_2}&h_{\mathrm{eg}}&0&\lambda\a{f_1}\\
		\lambda\adag{f_1}&0&h_{\mathrm{ge}}&\lambda\a{f_2}\\\hline
		0&\lambda\adag{f_1}&\lambda\adag{f_2}&h_{\mathrm{gg}}
	\end{array}\right),
\end{equation}
with $h_{xx'}=\omega_{x,1}+\omega_{x',2}+\dOmega$ for $x,x'\in\{\mathrm{e,g}\}$, with the various terms of the Hamiltonian being grouped accordingly. More compactly,
\begin{equation}\label{eq:tridiag}
H_{f_1,f_2}=\left(\begin{array}{c|c|c}
		h_{2}&\lambda A_{1,2}&0\\\hline
		\lambda A^\dag_{1,2}&h_{1}&\lambda A_{0,1}\\\hline
		0&\lambda A^\dag_{0,1}&h_{0}
	\end{array}\right),
\end{equation}
all terms to be found with direct inspection with Eq.~\eqref{eq:twoatom2}, bringing about a simple block tridiagonal structure; in particular, note that
\begin{equation}
A_{1,2}=\begin{pmatrix}
	a(f_2)&a(f_1)
\end{pmatrix},\qquad A_{0,1}=\begin{pmatrix}
a(f_1)\\a(f_2)
\end{pmatrix},
\end{equation}
We are now in a situation similar to that of the rotating-wave spin--boson model. When $f_1,f_2\in\hilb$, the Hamiltonian is self-adjoint by construction, with domain $\mathcal{D}(H_{f_1,f_2})=\mathbb{C}^4\otimes\fock\simeq\mathcal{D}^{(0)}\oplus\mathcal{D}^{(1)}\oplus\mathcal{D}^{(2)}$, where $\mathcal{D}^{(0)}=\mathcal{D}(\dOmega)=\mathcal{D}^{(2)}$ and $\mathcal{D}^{(1)}=\mathcal{D}(\dOmega)\oplus\mathcal{D}(\dOmega)$. When $f_1,f_2\in\hilb_{-1}\setminus\hilb$, by interpreting $\a{f_j},\adag{f_j}$ in the sense of Prop.~\ref{prop:af} we have
\begin{eqnarray}
    A_{1,2}:\hfrak^{(1)}_{+1}\rightarrow\hfrak^{(2)},\qquad A^\dag_{1,2}:\hfrak^{(2)}\rightarrow\hfrak^{(1)}_{-1};\\
    A_{0,1}:\hfrak^{(0)}_{+1}\rightarrow\hfrak^{(1)},\qquad A^\dag_{0,1}:\hfrak^{(1)}\rightarrow\hfrak^{(0)}_{-1},
\end{eqnarray}
suggesting that, by a suitable choice of domain generalizing Eq.~\eqref{eq:singdom}, it is possible to define a singular $2$-atom spin--boson Hamiltonian with form factors $f_1,f_2\in\hilb_{-1}$, that may be approximated in the norm resolvent sense by sequences of regular models. Similar considerations can be done for higher values of $N$, for which again a block tridiagonal structure analogous to Eq.~\eqref{eq:tridiag} emerges.

Finally, we point out that the scheme outlined above is sufficiently flexible to account for a more general choice of the operator $K$ in Eq.~\eqref{eq:freesbn}---that is, to include a suitable class of \textit{spin--spin interactions}. Indeed, the procedure can be reproduced without substantial effort by just requiring $K$ to be block-diagonal with respect to the decomposition~\eqref{eq:decomp} of the Hilbert space, that is,
\begin{equation}
    K\hfrak^{(j)}\subseteq \hfrak^{(j)}
\end{equation}
for all $j=0,\dots,N$. In words, spin--spin interactions that preserve the total number of atoms in their excited state can be included in this scheme as well. As a concrete example, in the case $N=2$ outlined above we have the freedom to add an interaction term which flips the states of the two spins whenever they are different: such a term will merely add off-diagonal terms to the central block of the Hamiltonian $H_{f_1,f_2}$ in Eq.~\eqref{eq:twoatom2}, but its block tridiagonal structure represented in Eq.~\eqref{eq:tridiag}, up to a proper redefinition of $h_1$, will not be disrupted. 

\section{Perspective: interior-boundary conditions for spin--boson models}\label{sec:ibc}

The results of~\cite{gsb}, together with the partial results presented in this paper, represent a first attempt towards a more systematic theory of GSB models with singular form factors. In this regard, it is useful to point out that such form factors---typically, exhibiting ultraviolet (UV) divergencies---have been long encountered in various other models of nonrelativistic quantum mechanical matter interacting with boson fields.

In these cases, one typically finds a scenario analogous to the one for GSB models: for sufficiently mild divergencies one seeks a form-theoretical approach; when such techniques are not feasible, a renormalization procedure analogous to the one considered by Nelson for the eponymous model~\cite{nelson1964interaction} may be attempted---again, without generally retaining information on the domain. In the latter case, the limiting Hamiltonian, say $H_\infty$, is obtained as the norm resolvent limit of $H_\Lambda+E_\Lambda$ as $\Lambda\to+\infty$, where $H_{\Lambda}$ is a properly regularized Hamiltonian (e.g. via a cutoff procedure) and $E_\Lambda$ a diverging energy shift. This procedure shows similarities with the renormalization of the spin--boson model for $f\in\hilb_{-2}$ presented in Section~\ref{sec:sb}: however, roughly speaking, in that case the excitation energy of the qubit ``takes charge'' of the diverging energy shift instead of the Hamiltonian itself.

A novel approach to UV divergencies in quantum field theories which, differently from the techniques briefly summarized above, can be adapted to different scenarios and accounts for an explicit determination of the operator domain, was recently proposed by Teufel and Tumulka~\cite{teufel2016avoiding,teufel2021hamiltonians} and has been successfully applied to diverse other particle‐-field Hamiltonians, see e.g.~\cite{lampart2018particle,lampart2019nelson} (also cf.~\cite{binz2021abstract} for the abstract setting). This approach is based on a systematic use of \textit{interior-boundary conditions} (IBC), that is, abstract boundary conditions relating sectors with distinct number of field excitations. 

The basic idea at the core of this approach is to employ a restriction--extension procedure analogous to the one used to define singular perturbations of possibly infinite rank~\cite{posilicano2008extensions} (also cf.~\cite{posilicano2020self}). Without entering into details, the procedure can be roughly summarized as follows:
\begin{itemize}
\item starting from the decoupled (self-adjoint) Hamiltonian $H_0$, one considers the restriction of $H_0$ to a properly chosen dense subspace of the Hilbert space, thus obtaining a ``minimal'' operator admitting infinitely many self-adjoint extensions;
\item such self-adjoint extensions are then characterized via suitable abstract boundary conditions on the domain of the adjoint of the minimal operator;
\item one finally searches for the boundary conditions corresponding to the formal, UV-divergent operator---i.e. such that, by applying the formal expression to the states satisfying said boundary conditions, all divergent terms cancel out.
\end{itemize}
In the process, the domain of the adjoint is shown to comprise vectors with a ``regular'' component in the domain of the decoupled Hamiltonian $H_0$, plus an additional ``singular'' part. As such, the procedure sketched above clearly shows important similarities with the renormalization procedure of the models discussed in Sections~\ref{sec:sb}--\ref{sec:multi}. In both cases, the key idea is to properly redefine the domain of the decoupled Hamiltonian by means of a singular term chosen so that the divergencies produced by the formal Hamiltonian cancel out. 

While leaving a more detailed discussion to future works, the IBC method can be already predicted to be consistent with the results of~\cite{gsb}. Indeed, applying the same steps followed e.g. in~\cite[Section 1]{lampart2019nelson}, the self-adjointness domain of the singular spin--boson model of Section~\ref{sec:sb} ends up being the space of vectors $\Psi\simeq\Psie\oplus\Psig$ such that the vector
\begin{eqnarray}\label{eq:ibc}
\Psi+\frac{1}{H_0+1}\sigma_-\otimes\adag{f}\Psi\simeq\begin{pmatrix}\Psie\\\Psig\end{pmatrix}+\lambda\begin{pmatrix}
0\\\frac{1}{\dOmega+1}\adag{f}\Psie
\end{pmatrix}
\end{eqnarray}
(with $\sigma_-$ as in Eq.~\eqref{eq:ladder}) is in $\mathcal{D}(H_0)=\mathcal{D}(\dOmega)\oplus\mathcal{D}(\dOmega)$, which is achieved by taking $\Psie\equiv\Phie\in\mathcal{D}(\dOmega)$ and $\Psig=\Phig-\lambda(\dOmega+1)^{-1}\adag{f}\Phie$ for some $\Phig\in\mathcal{D}(\dOmega)$, thus obtaining the very same expression as in Eq.~\eqref{eq:singdom}.

This argument strongly supports the IBC method as a natural candidate for a general theory of GSB models with singular form factors---even those not sharing the rotating-wave structure of the Hamiltonians discussed in Sections~\ref{sec:sb}--\ref{sec:multi}. Of course, while the general recipe to construct such models will be the same, the results will crucially depend on the particular choice of the operators mediating the matter--field interaction: for instance, replacing the operator $\sigma_-$ in Eq.~\eqref{eq:ibc} with another one will generally cause \textit{both} components of the wavefunctions in the domain to acquire a singular part, thus preventing an explicit evaluation of the resolvent.

\section{Concluding remarks}\label{sec:concl}
In this work we have reviewed the application of the formalism of singular annihilation and creation operators, defined as continuous maps on a scale of Fock spaces, to the nonperturbative study of simple models describing a rotating-wave interaction, mediated by non-normalizable form factors, between quantum systems with a finite number of degrees of freedom (e.g. spins) with a structured boson field. To this extent, we have summarized known results for the simplest instance of such systems, i.e. the rotating-wave spin--boson model, and discussed the application of similar techniques to more complicated systems. Future works shall be devoted to a more detailed developments of the ideas presented in the last sections.

As a common thread, for such Hamiltonians the accommodation of form factors in the class $\hilb_{-1}$ seems to only involve a careful choice of the operator domain, compatibly with existing perturbative results for all models in the class of generalized spin--boson models; the case $\hilb_{-2}$ is instead much less trivial, already in the case of the spin--boson model, where a nontrivial renormalization procedure is required.

\begin{acknowledgement}
We acknowledge support by MIUR via PRIN 2017 (Progetto di Ricerca di Interesse Nazionale), project QUSHIP (2017SRNBRK). This work is partially supported by Istituto Nazionale di Fisica Nucleare (INFN) through the project “QUANTUM” and by the Italian National Group of Mathematical Physics (GNFM-INdAM). We also thank Paolo Facchi for many fruitful discussions.
\end{acknowledgement}

\printbibliography

\end{document}